\documentclass[aps,prb,twocolumn,showpacs,floatfix,superscriptaddress]{revtex4-1}

\usepackage{booktabs} 
\usepackage{graphicx}
\begin{document}

\title{Introduction to Computational Thinking: a new high school curriculum using CodeWorld}

\author{Fernando Alegre}
\affiliation{Gordon A. Cain Center for STEM Literacy,
    Louisiana State University,
    Baton Rouge, LA 70803}
\author{John Underwoood}
\affiliation{School of Education,
    Louisiana State University,
    Baton Rouge, LA 70803}
\author{Juana Moreno}
\affiliation{Department of Physics \& Astronomy
    and Center for Computation \& Technology,
    Louisiana State University
    Baton Rouge, LA 70803}
\author{Mario Alegre}
\affiliation{Department of Physics,
  Pennsylvania State University,
  University Park, PA 16802}

\begin{abstract}
The Louisiana  Department of Education partnered with the Gordon A. Cain Center at LSU
to pilot a Computing High School Graduation Pathway. The first course in the
pathway, Introduction to
Computational Thinking (ICT), is designed to teach programming and reinforce
mathematical practice skills of nine-grade students,
with an emphasis on promoting higher order thinking.
In 2017-18, about 200 students and five teachers participated in the pilot, in 2018-2019 the participation increased to 400 students, and in the current 2019-2020 year about 800 students in 11 schools are involved.
Professional development starts with a five-week intensive summer institute,
which is complemented with follow-up Saturday sessions and coaching support
during the academic year.
After describing the course content and briefly the teacher training, we discuss the
data we have collected in the last two years.
The overall student reception of the course has been positive,
but the course was categorized by most students as
hard. However, the Computing Attitude Survey analysis indicates that the difficulty of the course did not demotivate the students.
The pre-post test content assessments show that students learned not
only the language, but also general principles of programming, logic and modeling,
as well as use of variables, expressions and functions.
Lessons learned during the pilot phase motivated changes, such as emphasizing
during PD the need to provide timely feedback to students, provide detailed rubrics for the projects and reorganize the
lessons to increase the initial engagement with the material.
After two years of running pilots, the course is becoming student-centered, where most of the code and image samples provided in the lessons are based on code
created by previous students.
\end{abstract}


\maketitle

\section{Introduction}
\label{sec:overview}

Our project started in 2015, when we were contacted by the East Baton Rouge Parish School System (EBRPSS)
to help develop computer science curricula for a new STEM magnet
high school, 
to offer new opportunities to the under-served population of the
district, which consists of 85\%
minority and 75\% economically disadvantaged students. 
We were tasked with creating the curriculum, including its assessment
and the delivery of the summer teacher training. Additionally, the curriculum had to be designed in such a way that teachers
of other academic subjects could quickly learn it, since there were no computer science teachers available in the area. 
The first course in this set is Introduction to Computational Thinking (ICT), an introductory
programming course offered to eighth or ninth graders who are concurrently taking an Algebra I.
The course teaches the conceptual foundations of
coding in a language syntax and semantics that follow closely the language of algebra.
It is not intended to be a math remediation course, but rather to highlight the connections to algebra, geometry and science modeling.

During the 2016-2017
academic year, we conducted several 3-month pilot tests of the course and developed an assessment
instrument, the \emph{Conceptual Foundations of Coding Test}, which was
vetted with about 100 students. A full-course pilot was deployed in the 2017-2018 academic
year. At that time,
the Louisiana Department of Education
(DoE) became interested in the ICT curriculum and partnered with the Cain Center
to create and pilot a Computing High School Graduation Pathway,
following the model pioneered by the EBRPSS STEM magnet high school.
The Pathway offers a hybrid curriculum that prepares students both for college and to enter
the workforce after graduation.

During 2017-2018, the course was taken by more than 200 ninth grade students in four different schools.
Approximately 400 students in ten schools in
eight school districts were enrolled in ICT for the 2018-2019 academic year,
and there are 800 students enrolled in the 2019-2020 academic year.

In the summer of 2017, we conducted our first Professional Development program, which is an intensive five-week
professional development summer institute. In 2017 we trained eight teachers, with an additional nine teachers in 2018, and most recently nine more teachers. The teachers were absolute novices with respect to programming. They were placed into student roles as the first part of
their training, where they completed all the programming assignments, presented them to their peers, and practiced modifying their code according to
the feedback received. The teachers were instructed in pedagogical techniques and lesson design throughout the summer. At the end of the summer PD, the teachers felt comfortable
enough to teach the course and to modify the assignments to meet their school's unique cultures and needs.
The majority of the teachers participating were certified in
either secondary math or science, but some were certified in other instructional areas. For example, in the 2017 training there was one social studies and one career and technical education teacher. In
the 2018 training there were two social studies teachers, and in the 2019 training there two social studies and two computer science teachers.

All the activities are programmed in CodeWorld\cite{codeworld},
a web-based integrated development environment which was initially designed for middle school students,
that uses a simplified variant of the Haskell language.
The lessons are organized in
units that follow the concepts of Computational Thinking, with the syntax of the
language being presented at the beginning of each semester. 
However, very little emphasis is placed on
teaching the language, whose features are introduced only when needed.
In the first semester, only expressions, variables and functions are used. No conditionals,
looping constructs or data structures are needed for the programming
assignments. In the second semester, lists and tuples are the only new syntactic features needed, and looping constructs are based on a second-order function, called \verb|foreach|,
which is a regular function with no special syntax.

\begin{table*}[tb]
\begin{center}
  \caption{ICT Course outline.}
  {\begin{small}
\begin{tabular} {|p{1.in}|p{5.5in}|}
\hline \hline
  Unit & Content \\
  \hline \hline
  The Software Development Cycle &
  Students learn how to use an IDE, how to draw basic shapes, how to overlay several pictures and move them around the screen. They also learn about design techniques, such as creating prototypes and using pseudo-code to plan a program, and practice collaboration with pair programming and a collaborative creation of a scene, where each team member is in charge of a character or prop.
  \\
\hline
Abstraction and Decomposition &
Students learn to map expressions to syntax trees, handling order of operations, and using trees to represent other aspects of code, such as dependencies between variables and organization of layout into nested layers. They also learn how to use an object dimension as a unit of measurement for other objects (e.g., 2.5 smileys wide, 3 monsters high)
and how to combine rotations, translations and scalings to create complex mosaics or quilt patterns. \\
\hline
Patterns and Regularity &
Students use repetition to create regular polygons, regular stars and create recursive patterns. They also learn about generating random patterns and irregular grids, and use them to generate a procedural map of a neighborhood.  \\
\hline
Data and Calculations & 
Students learn to process lists to create bar charts and pie charts from scratch, create itemized bills including taxes and discounts, calculate weighted averages and compute areas of complex
settings, such as the area occupied by chairs and tables in a dining hall. \\
\hline
Models in Space and Time &
Students create simple games (rock,paper,scissors; dice rolling games; tic-tac-toe) and simple animations (characters performing repetitive circular or linear motion; see-saws; slide shows; marquee messages) \\
  \hline\hline
\end{tabular}
    \end{small}
    }
\label{tab:ICTCurriculum} 
\end{center}
\end{table*}

\section{Development Process}
\label{sec:research}

\subsection{Foundational Stages}

The discipline of computer science currently has a large, if not traditionally recognized, impact on  many other high school core subjects,and therefore it should not be studied
in isolation~\cite{heintz16}. Computer science must also have the integration of  skills and content that reflect the real word connections it has to math and science~\cite{bart14}. Learning computer science is not about learning a specific programming language. It is also not simply learning commands and techniques on how to program. 
The term \emph{Computational Thinking} (CT) was introduced in education to
convey this type of reasoning
\cite{weintrop16CTMathSc,angeli16CT-K6,voogt15CTcore,
grover2015designing, grover13ctk12,wing2006computational}.
Computational Thinking is not confined to strictly  programming, but it is within the programming environment where CT manifests itself most prominently. It can be difficult to understand CT fully without exposure to programming.
\cite{grover13ctk12,denning17CT}.

In our training materials, we try to provide insight about the meaning
of computational thinking, so we explain to the teachers that
when someone is thinking computationally, in our view, they do the following:
\begin{enumerate}
    \item Use introspection to observe their own thought process as if it were performed by a machine and express their thoughts explicitly and without any ambiguity.

    \item Imagine in their head a computer running a given program and anticipate the outcome without actually running the program.

    \item Reason constructively, as the purpose of computing is to construct a solution. Computing works under a closed world assumption, where only those entities explicitly built are assumed to exist.

    \item Invent a process to solve a problem as a series of mechanical steps, where each step requires no intelligence to perform. The intelligence contained in a program is an emergent feature and cannot be pinpointed to any particular line in the code.

    \item Think in terms of causality. A function is not just a relationship between an input and an output, as it would be in mathematics. It is also a process that causes the computer to produce an output when the given input is consumed. This process occurs in time, and so the input must exist before the output can exist. Computations change the world.
    
    \item Reason by proxy: Distinguish between what a concept is and how it is represented. For example, represent a polygon as a list of pairs of coordinates.

    \item Establish relationships between concepts by writing equations between their corresponding representations. For example, move a polygon horizontally by adding the same number to each X coordinate in the corresponding list.

\end{enumerate}

Computational thinking is about expressing thoughts in a formal system, in a way that is actionable by an automated system. Programming languages are not the only possible formal systems in which computational thinking can be expressed, but they are the most accessible and prone to automation. Thus, using programming as a vehicle for computational thinking is a natural choice. 

Unfortunately, in many
elementary and middle school settings, the term CT has become synonymous with
either \emph{computing with no programming} or \emph{block-based programming}.
This interpretation omits the central tenet of Computational Thinking, which
is the  \emph{building of high-level abstractions that can be executed by a computer}
\cite{wing08compthink}.
Currently, there is a need to have a high school course that introduces
CT within the context of substantial amounts of programming
with clear connections to math and science. This CT course should depend as little
as possible on the extensive knowledge of a particular language or technology. The CT course would be a natural progression for students to take along with Exploring Computer Science (ECS)  and
Computer Science Principles (CSP).

For the most part, ECS, CSP and block-based programming courses rely
on the teachers to act as facilitators of instruction provided by an online system. This instructional model is based on the idea that students will 
learn even if they are not being directly instructed by their teacher.
However, a flaw of the model is the fact that many concepts, such as abstraction, are only developed through higher-order learning~\cite{feathers19,ramineni18}.

A recognizable factor for why they are on the rise is due to a current scarcity of teachers who know how to program. 
Students need direct interaction
with a teacher to master higher order thinking concepts. There is great value in having a teacher who can evaluate the students work, reflect on the student's progress, offer guidance to the student on ways that they may correct habits, and examine unique work products from the perspective of the student's intellectual evolution. These attributes have proven difficult to
evaluate effectively in an automated way.  

\subsection{Conceptual framework}

ICT is an elective course in Louisiana, where a majority of the students have historically demonstrated weak mathematical
skills for all grade levels. In designing the course, additional attention was given to ways to help
students improve in their math skills as they learn computer science. This course was not intended to be strictly a math remediation or math intervention course, but rather an integrated part of a STEM elective pathway. The learning objectives were established and designed to be recurring throughout each of the units. The learning objectives are not isolated to specific lessons.

The learning objectives include:

\begin{itemize}
\item Develop a procedural understanding of the pillars of
Computational Thinking: recognize patterns and regularity,
decompose problems into smaller problems,
formulate and solve simplified problems, generalize solutions and encapsulate solutions.
\item Acquire experience with algebraic manipulation of complex expressions
\item Use mathematical functions to model artifacts, such as diagrams or animations.
\item Transform many data items as if they were a single entity
\item Organize data hierarchically
\item Calculate totals, averages and quantities using rates, such as taxes
and discounts.
\item Use random sampling to explore instances of relationships and find the general case
\end{itemize}

Our approach is inspired by the work of \cite{felleisen04SchemeCS}
and Bootstrap~\cite{Schanzer13,schanzer15transfer}, due to their
promising results concerning transfer between programming and mathematics
~\cite{Schanzer13,schanzer15transfer}. They introduced the \emph{design recipe}, which is a series of steps for guiding students when they are trying to create a function: write a definition in English, then describe the inputs and outputs, then provide at least 3 examples, then look at what is common in those examples (the template) and what changes from example to example (the variables), and finally give names to those variables.

However, we differ from Bootstrap in several different ways. We have developed
a full-year curriculum centered on CT instead of a 17-hour intervention focused on math word problems. Our use of Haskell makes writing function definitions very lightweight, so students are encouraged to create lots of functions. Also, the lazy evaluation model relieves us from the need to have special syntax for program control. We have also extended the design recipe with the introduction of random variables, so that students create random samples of uses of a function after (or instead of) providing examples with fixed numbers. Finally, we put more emphasis on modeling techniques and using the software development cycle rather than on guided exercises based on code templates.

\begin{figure}[h]
\centerline{\includegraphics[width=0.20\textwidth]{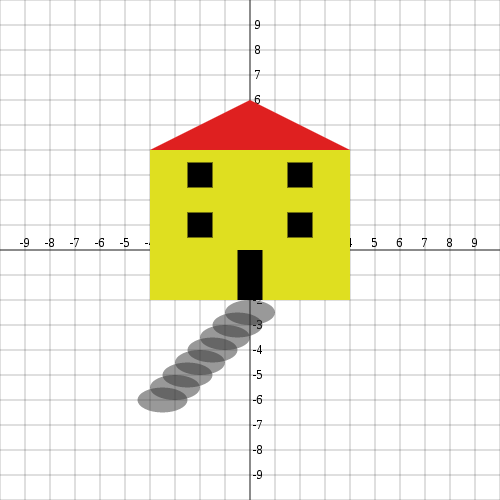}}
\caption{\label{fig:house}A drawing of a house}
\end{figure}

\subsection{CodeWorld activities}

All the activities are programmed in CodeWorld\cite{codeworld},
which uses a very
limited set of graphical primitives to draw circles, rectangles, and text. It is then possible to apply translations, rotations, scalings and colors to them. Smaller
elements can be combined into more complex shapes via the overlay operator
(denoted by \&).  Animations are represented as functions that depend on
a parameter, namely the time in seconds since the animation started. The
language follows a syntax very similar to mathematical notation, and the
evaluation semantics follows exactly the same rules as algebra.

Here is a complete CodeWorld program to draw a house (see Fig.~\ref{fig:house}):

\begin{verbatim}
program  = drawingOf(house(red,yellow)
         & coordinatePlane)
house(rcolor,fcolor) =
  colored(roof,rcolor)
  & windows & door 
  & colored(facade,fcolor)
  & pathway
roof     = solidPolygon([ (-4,4),(4,4),(0,6) ])
windows  = floor2 & floor3
floor2   = translated(window,-2,1)
         & translated(window,2,1)
floor3   = translated(floor2,0,2)
window   = solidRectangle(1,1)
door     = translated(solidRectangle(1,2),0,-1)
facade   = translated(solidRectangle(8,6),0,1)
pathway  = overlays(tile,8)
tile(n) = translated(stone,-(n-1)/2,-1.5-(n+1)/2)
stone = colored(oval,translucent(grey(0.2)))
oval = scaled(solidCircle(0.5),2,1)
\end{verbatim}

Practically all the syntax of the language is illustrated in the previous
program, and all programs are written in exactly the same format (a list of
lines that read \verb|head = body|) with \verb|program| being the starting
point of the execution of the program.  Functions are defined in the same way
as variables, but the head includes parameters.  No special constructions
for loops or conditionals are necessary. Definite loops are provided by
library functions, such as \verb|overlays|, which works as follows:
the expression \verb|overlays(f,n)| is equivalent to
\verb|f(1) & f(2) & ... & f(n)|.
Indefinite loops are created by recursive definitions. Conditionals
are produced by having functions with special cases, which are created by
adding a vertical bar and a condition to their definition. For example,
the absolute value would be defined by the following two lines:

\begin{verbatim}
absoluteValue(x) | x <  0  = -x
absoluteValue(x) | x >= 0  =  x
\end{verbatim}

In the second semester, lists and tuples are introduced. Basic list usage
needs 3 additional symbols: \verb|[|, \verb|]| and \verb|#|, to build
a literal list, and to access the n-th element, respectively.

The simplified version of Haskell we are using stops here. No advanced
features of the language (such as typeclasses, IO or monads) are exposed
to students. In a sense, our use of Haskell provides the same affordances that a block-based
language would, because the key features of block-based languages are their
simple, barebones syntax, as opposed to regular programming languages, and the
avoidance of errors due to misspelling or misuse of variables and constructs
placed in the wrong spot~\cite{Weintrop15BlockProg,weintrop15CommAss}. Haskell
shares both of these features, because in addition to the simple syntax
explained above, the advanced type-inference features of Haskell catch
practically all misspellings and misuses of variables and functions.

\subsection{Curriculum content}
\label{sec:content}

The ICT curriculum comprises five units, where the first two units take approximately eight weeks each, and the last three units take approximately
five weeks each. In the 2019-2020 version of the curriculum, all the activities in the first four units include samples created by students who took the course in the previous two years.
See Fig. 2 for the samples in an activity where students are asked to create a diagram of a cell.

Assessment is based on a project at the end of each unit, plus a midterm project and a final project. Our team developed analytical rubrics that present the criteria and levels of performance for each assignment. The rubrics are tiered from minimal, to lower, to mid-level, and finally high attainment. Each tier contained descriptors with point values. Each attribute was aligned to the learning objectives, which are stated at the start of each lesson and integrated in the activities that build to the project. For example, in a project to create an analog clock face, students were evaluated on the following: whether they used expressions with variables (high score) or magic numbers (low score); whether they repeated the code for the hour hand and the minute hand (low score) or created a function to handle both (high score); whether they created different nested layers for the hour ticks, the minute ticks and other elements (high score) or they had a flat layout (low score); whether their code printed redundant elements, such as printing 12 o'clock twice (low score) or handled ranges properly, including only one end (high score); whether they used local variables (high score) or only global variables (low score); whether they followed good practices when naming, indenting and grouping parts of the code (high score) or not (low score); whether the calculations to convert hours and minutes to degrees of rotation were correct (high score) or not (low score); and whether the output showed an analog clock with hour ticks, minute ticks and numbers at each hour (high score) or only some of those elements (low score). Finally, extra credit was given for creativity in their aesthetic design.

Table \ref{tab:ICTCurriculum} lists the title and a brief summary of the
content of each unit.
  
\begin{figure*}[tb]
  \minipage{0.16\textwidth}
  \includegraphics[width=\linewidth]{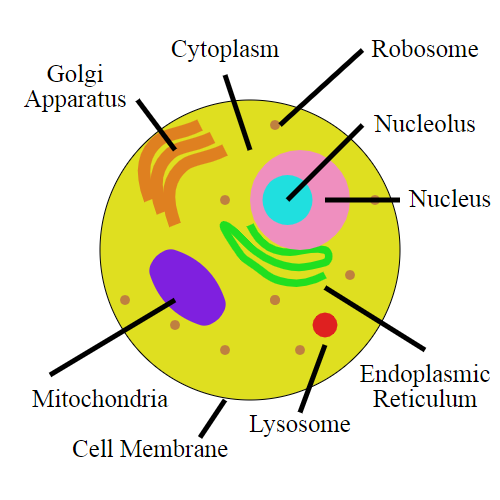}
  \endminipage\hfill
    \minipage{0.16\textwidth}
  \includegraphics[width=\linewidth]{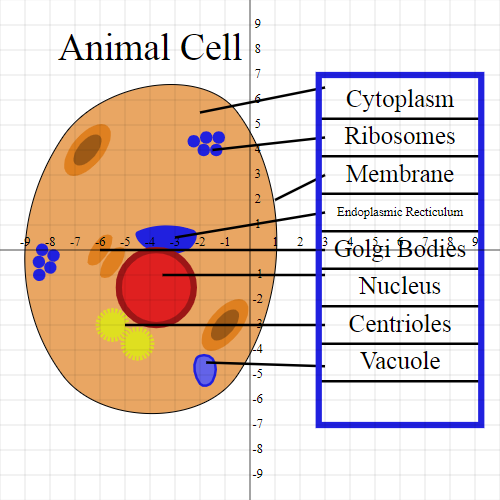}
  \endminipage\hfill
    \minipage{0.16\textwidth}
  \includegraphics[width=\linewidth]{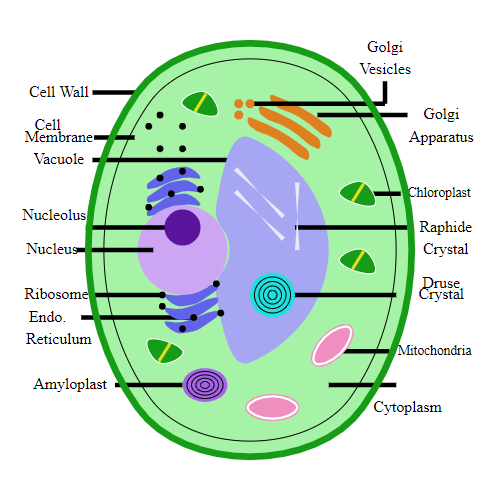}
  \endminipage\hfill
  \minipage{0.16\textwidth}
  \includegraphics[width=\linewidth]{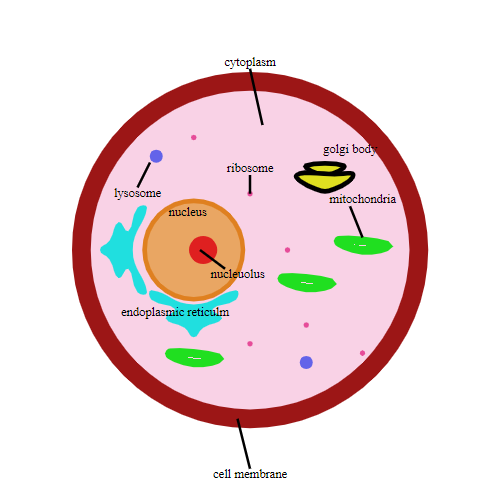}
  \endminipage\hfill
    \minipage{0.16\textwidth}
  \includegraphics[width=\linewidth]{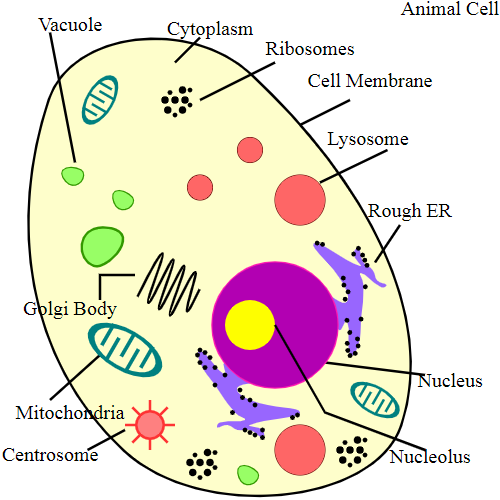}
  \endminipage\hfill
    \minipage{0.16\textwidth}
  \includegraphics[width=\linewidth]{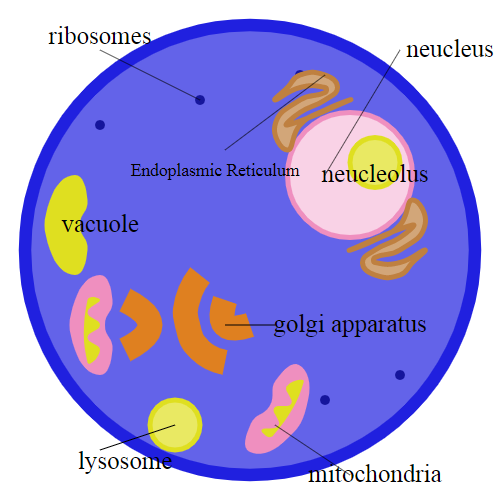}
  \endminipage
  \caption{Six examples of cells designed by different students in the 2017-2018 academic year.}
  \end{figure*}
  
  \begin{figure*}[tb]
  \minipage{0.24\textwidth}
  \includegraphics[width=\linewidth]{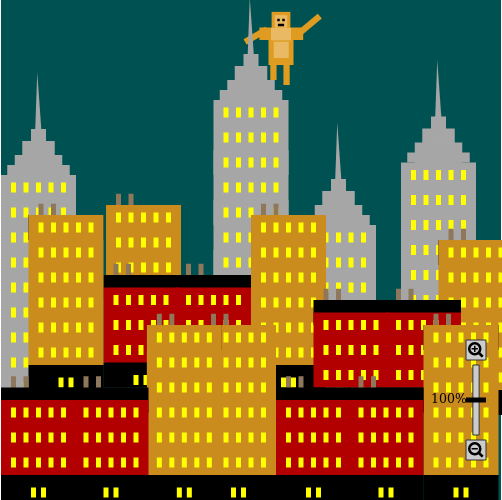}
  \endminipage\hfill
    \minipage{0.24\textwidth}
  \includegraphics[width=\linewidth]{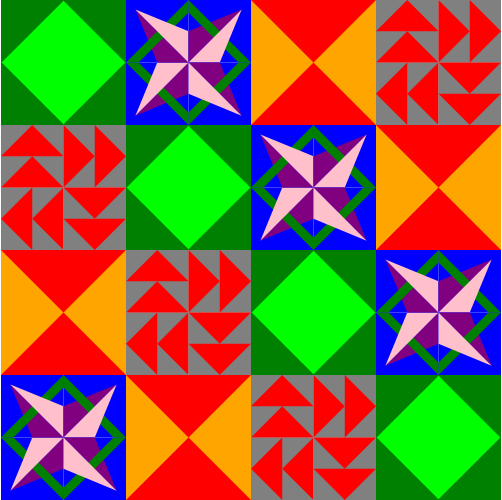}
  \endminipage\hfill
  \minipage{0.24\textwidth}
  \includegraphics[width=\linewidth]{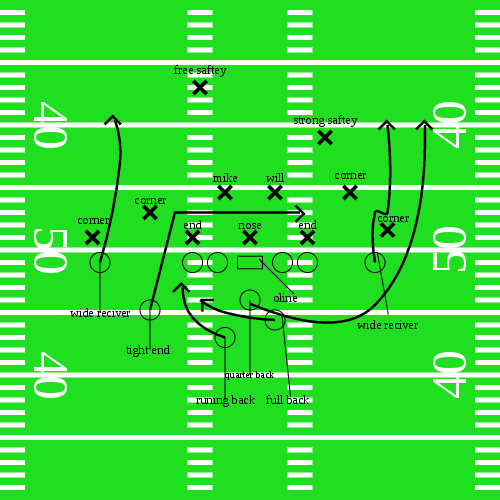}
  \endminipage\hfill
    \minipage{0.24\textwidth}
  \includegraphics[width=\linewidth]{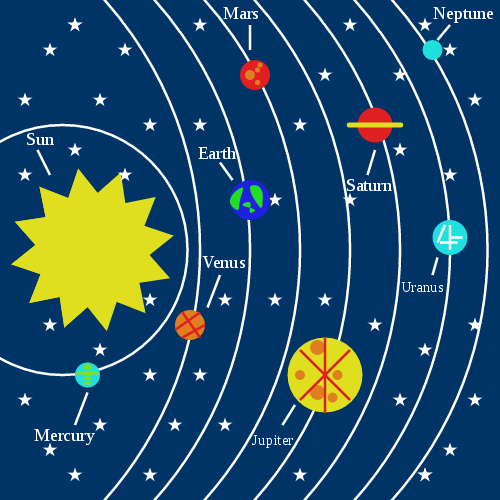}
  \endminipage
  \caption{Examples of student made projects for different units.}
\end{figure*}

\subsection{Technical considerations}

Our choice of programming environment was also influenced by the following properties:
1) The programming language should make
it easy for students to build high level abstractions;
2) The language should also have a syntax and semantics
as similar to algebra as possible;
3) No prior or additional knowledge of coding or software should be needed
by teachers to produce code for the lessons;
4) Execution of any component of the system should not depend on
any third party service or product;
and 5) The programming paradigm should preferably be functional.

One technical restriction on our choice of programming
environment was due to the fact that many Louisiana schools have policies concerning which software can be run on
their computers. Often the computing environment is optimized for
use on standardized testing platforms, which can prevent root access and local
installation of software. In addition, policies in many schools prevent students from being required to register with third-party organizations and to submit student work to external web sites. Given these conditions, we elected to use
a Web-based environment that required no local installation and could be used without restrictions and without the need for students to register or provide any personal information.
The next requirement our team faced was that we were required by the Louisiana Department of
Education,
which partially funded our project, to rely upon fully open source software. Our final requirement was that as a team we wanted to do graphics-based programs rather than text-based programs.
Given all of the aforementioned requirements the number of possibilities we considered was limited. For example, at the project's onset there was no fully open source, fully online version of Python for graphics programming.

\section{Related and future work}
\label{sec:related}

The idea of using coding to help students learn mathematics and science has a
long history.  Early attempts to  use coding as a tool were based on unguided
discovery~\cite{Papert80mindstorms}. This approach proved to be ineffective
for transfer \cite{mayer04strikesrule}.  Over the years, it has become clear
that transfer between programming and mathematics is difficult to initiate,
and whether it occurs or not depends strongly on the teaching  methodology
used~\cite{Pea83Logo,kurland86proghighschool,butterfield89techtransfer}.
Recent attempts to establish the link between programming and
mathematics have been based on a modern framework of computational
thinking~\cite{wing2006computational,wing08compthink,grover13ctk12}
and supported by modern theories such as convergent
cognition~\cite{rich13convergent}.

One of the few cases in which a project targeted the learning of mathematics
with coding and showed promising results is
Bootstrap, a 17-hour curriculum designed to be used either standalone
or embedded in a computer science or mathematics course.  It is one of
the few documented instances of transfer between programming and algebra.
\cite{Schanzer13,schanzer15transfer} attribute the favorable results of their
intervention to their use of a functional language as the medium and to the
absence of distracting features.

Since our program also features a functional language and absence of distracting features, we will investigate in future publications the capacity of our program to promote transference between mathematics and computational thinking.
As the ICT course is continuing to be adopted across Louisiana, a growing number of students will have taken ICT
in eighth or ninth grade. With an increase in student participants our team will be able to study the impact of functional programming in the learning of math and science.
Furthermore, as more math and science teachers are trained to teach ICT, we will be able to incorporate coding activities as  essential components in math and science courses.
Once complete, these activities will be accessible from a
virtual bank of math and science programming activities. The goal will be for these activities
to be adapted and incorporated by the teachers into their regular lessons that they currently do with their students.

Using this population of students and teachers trained in our ICT
course we can actually focus on the important question of whether students who know functional programming and
have access to instructional material that requires use of their programming
skills can reach a deeper understanding of math and science than students
who do not have access to that kind of instruction. In this future work comparable control populations will be attainable from local area schools that have yet to be trained.

\section{Impact}

\begin{figure} 
  \includegraphics[width=0.45\textwidth]{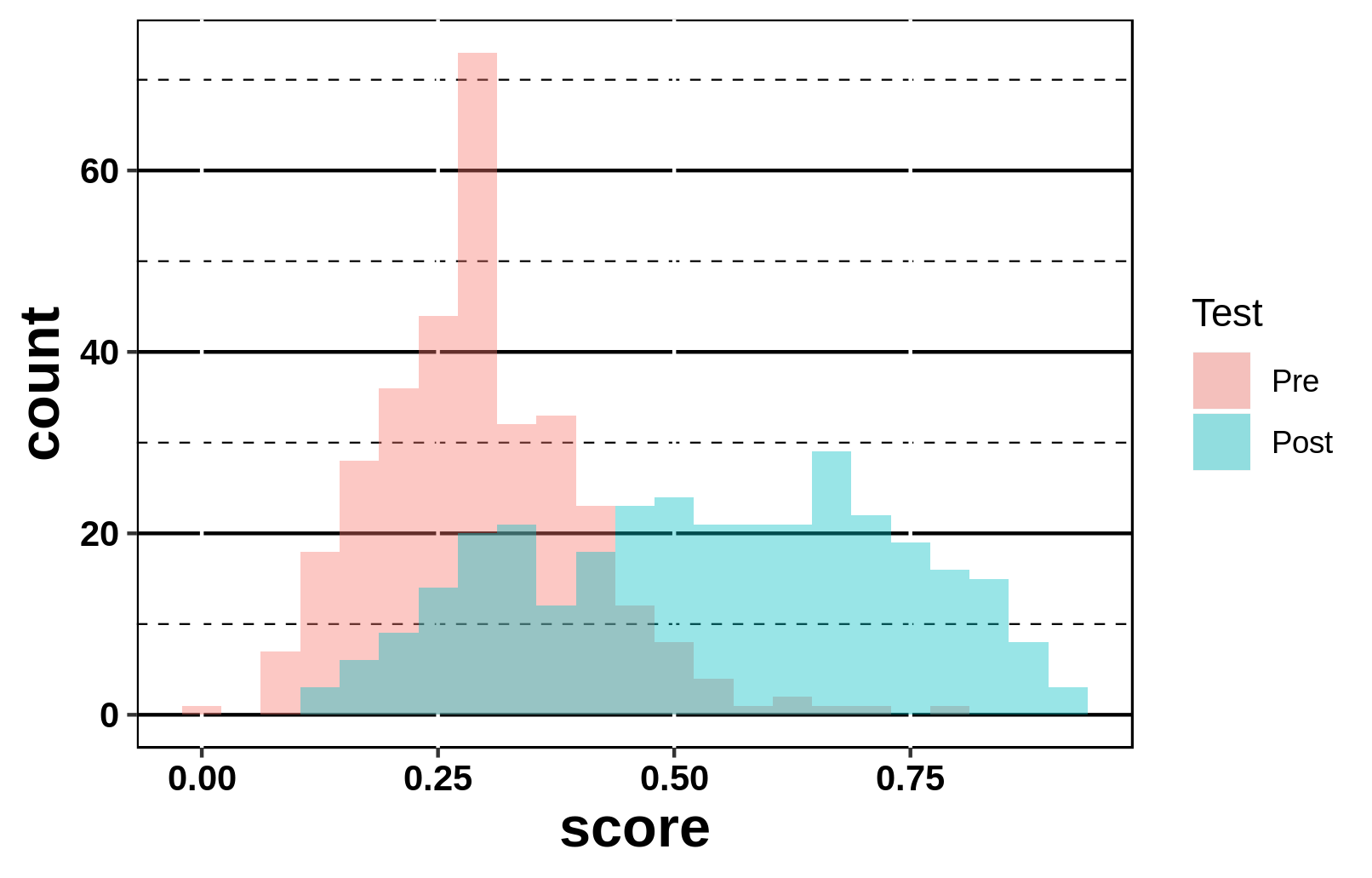}
  \caption{\label{fig:ICT}
    Histogram comparing the results of the pre- and
post-{\it Conceptual Foundations of Coding Test}.
}
\end{figure}

\begin{figure}
  \includegraphics[width=0.45\textwidth]{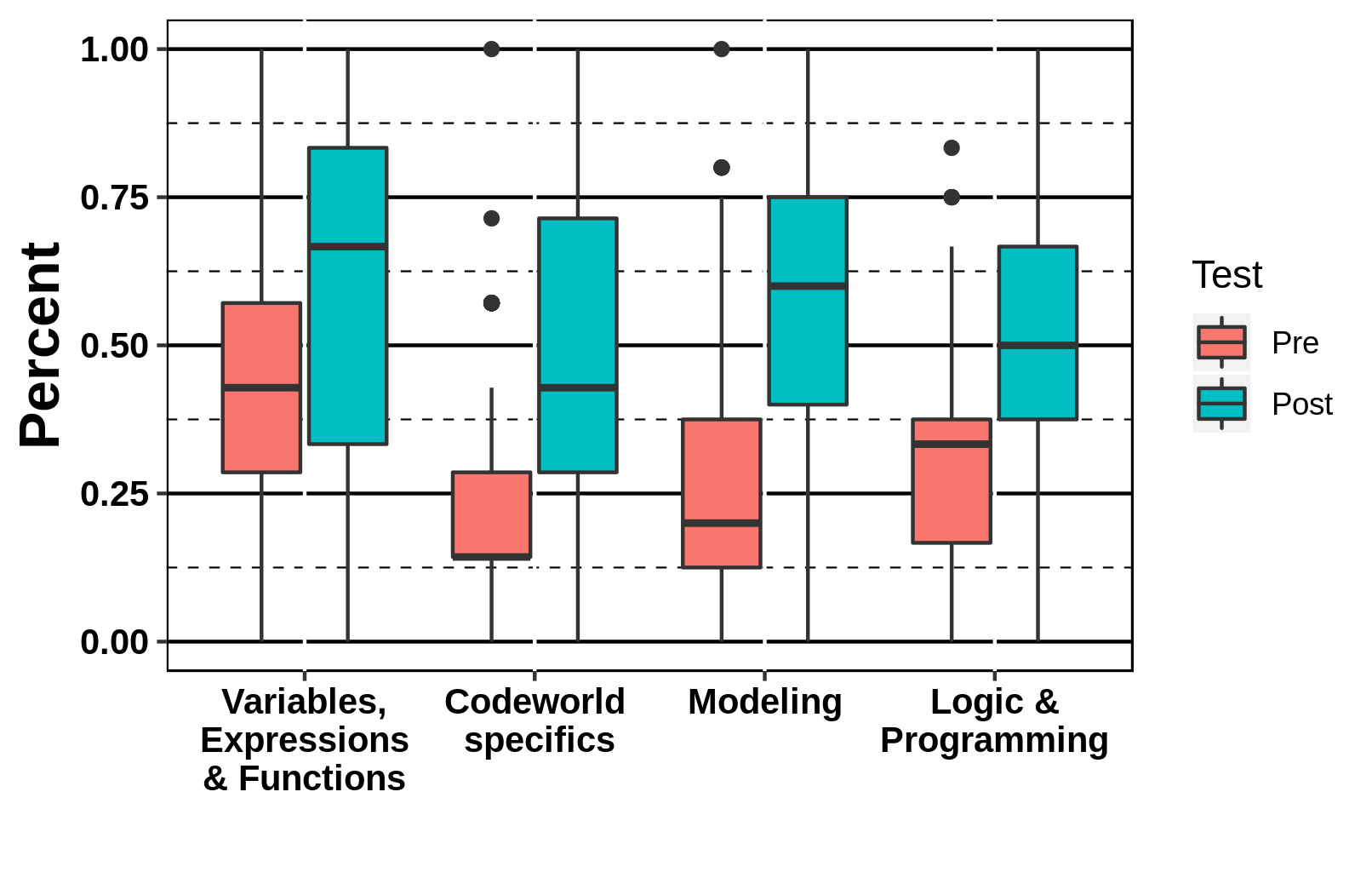}
  \caption{\label{fig:ICTFactors}
    Boxplots comparing results of the
{\it Conceptual Foundations of Coding Test} by categories:
  variables, expressions and functions; CodeWorld specifics; modeling;
and logic and programming.
  Boxes stretch from the 25th percentile to the 75th percentile of the
distribution.}
\end{figure}

\subsection{Results and Analysis}

The data presented here corresponds to academic years 2017-2018 and 2018-2019. In 2017-2018 the
course was deployed at four schools with five teachers and 208
students. Ten schools, 13 teachers and 395 students 
participated during the 2018-2019 academic year.

We collected pre- and post-{\it Conceptual Foundations of Coding Test}
results for 325 students.  The test captured an overall average growth of 24\% .
Fig.~\ref{fig:ICT} displays a
histogram comparing the pre- and
post-test results, showing that the distribution of the post-test moves to
higher scores and widens when
compared with the pre-test scores.  
The initial average score was $29.5 \pm 0.6\%$, and it increased
to $53.8 \pm 1.1\%$ 
at the end of the course.
A Wilcoxon signed rank test pairing pre- and post-results determined
that the difference between the distributions was statistically significant
($p < 2.2 \cdot 10^{-16}$), reinforcing that we can
distinguish the pre- and post-distributions of correct answers. 

Fig.~\ref{fig:ICTFactors} displays the pre- and
post-test results for the four categories
included in the test: variables, expressions and functions, \verb|CodeWorld|
specifics, mathematical modeling, and logic and programming.
The average of the post-results is higher than the average of the pre-results in all four categories.
The pre- and post distributions for all the categories are very significantly different,
with $p<10^{-12}$ as determined by the Wilcoxon signed rank test.

In the analysis of the Computing Attitudes Survey (CAS) data we followed the prescribed guidelines of Dorn and Tew~\cite{dorn15CAS} and 
analyzed the data utilizing the prescribed subcategories.
The data demonstrated that the student attitudes did not change after
completing the course, as can be determined by the fact that the shift in attitudes was minuscule ($0.02\pm0.02$), 
and the pre- and post-test results were very strongly correlated ($R=0.54$ and $p=1.3\cdot 10^{-10}$).

Looking at the correlations between the results of 
the Computing Attitudes Survey (CAS) and the gains in the
Conceptual Foundations of Coding Test (CFC), we find a positive correlation 
between the post results of CAS and the gains in CFC ($R=0.29$, $p=0.001$),
indicating that those students who had a more positive attitude at the end
of the course also tended to have the higher gains in learning. 

The overall reception of the course has been positive, but the course was
uniformly categorized by most students as hard. Our observations
indicate that students were not accustomed to having to use more than one
mathematical idea in a single problem. They were also unsettled by the fact
that the same image could be generated in many different ways, and there was
not a canonical correct way to write code for it. Nevertheless, the Computing
Attitude Survey analysis indicates that the difficulty of the course did not
demotivate the students. The pre-post test analysis
shows that the students learned not only the language, but also general principles of programming, logic and modeling,
as well as use of variables, expressions and functions.

\subsection{Conclusion}

While the need for teaching computational thinking is already well established, there is still controversy about whether programming should be included or not, or, as Denning~\cite{denning17CT} calls it, the clash between Traditional CT and New CT. Courses such as ECS or CSP are examples of New CT, but there is not much available in terms of courses that focus on Traditional CT. Due to its capacity for automation and formalization, programming is a natural vehicle for learning computational thinking. While Python and JavaScript courses are relatively available, they do not usually focus on CT. Instead, they follow traditional syntax-oriented approaches to teaching computing, with few connections to math and science. Those courses are more useful for students aspiring to be software developers than for the general student population. We have presented an alternative approach.

We have described the design and implementation of a secondary Computational Thinking course based on programming with connections to science and math. This course provides a proof of concept for curricula halfway between traditional programming language courses and recent computational thinking courses with limited programming content. This course addresses the need for computational thinking courses intended not only for future software developers but for all students no matter what they do later in their lives. We find that a focus on programming content does not need to be discouraging to students. Our approach is highly student-centered, and has been proven to be suitable for traditionally underserved populations. We also build on Bootstrap ideas and techniques and have opened a way to investigate many interesting connections between the learning of programming and the learning of mathematics and science, and we are excited to delve into them.

%

\end{document}